\documentclass[twocolumn,showpacs,preprintnumbers,prb]{revtex4}
\usepackage{times}
\newcommand{\figwidth}{3.25in}
\usepackage{graphicx}
\usepackage{amssymb}

\begin{document}

\title{On the use of  Purcell factors for plasmon antennas}
\author{A. Femius Koenderink$^*$}
\address{Center for Nanophotonics, FOM Institute for Atomic and
Molecular Physics (AMOLF), Sciencepark 104, NL-1098XG Amsterdam, The
Netherlands.  $^*$Corresponding author: f.koenderink@amolf.nl}
\begin{abstract}
The Purcell factor is the standard figure of merit for spontaneous
emission enhancement in microcavities, that has also been proposed
to describe emission enhancements for plasmonic resonances. A
comparison is made of quality factor, mode volume and Purcell factor
for single and coupled plasmon spheres to exact calculations of
emission rates. The paper explains why the Purcell factor is not
appropriate for plasmon antennas.
\end{abstract}


\maketitle

It is a paradigm of quantum optics that the internal dynamics of a
quantum system  can be controlled by placing the quantum system in a
 photonic environment that is resonant with radiative transitions of the source~\cite{Novotny}.
 Indeed, according to Fermi's Golden Rule, the
spontaneous emission decay rate is proportional to the density of
photon states that the photonic environment offers for spontaneous
decay~\cite{Novotny}. A resonant microcavity with a high quality
factor $Q$, and a mode confined in an ultrasmall volume $V$ enhances
the density of photon states by a factor know as the Purcell factor
$Q/V$~\cite{Novotny,cavityreview,Gerard}. Creating microcavities
with the highest $Q/V$ is a central target in solid state quantum
optics, with success being reported with photonic crystals,
micropillars, and SiO$_2$ microspheres and
microtoroids~\cite{cavityreview,Gerard}.

 Recent advances in
nanofabrication have provided new impetus to the field of
plasmonics, that creates highly confined electromagnetic resonances
by using the response of free electrons in
metals\cite{plasmonreview}. In particular, there are widespread
activities that exploit plasmon antennas for enhanced fluorescence
and quantum optics with single
plasmons~\cite{plasmonreview,anger,sandoghdar,mertens,jaime,taminiau,moerner,lukin}.
Due to absorption and radiative loss, the quality factor of
plasmonic devices is extremely low ($Q<10^2$). Nonetheless, the
truly nanoscale confinement leads to estimates for the Purcell
factor of order $F= 10^3$, competitive with dielectric
cavities~\cite{lukin,polman,maierV}.  Motivated by the tremendous
success of the Purcell factor  as a quantitative figure of merit for
dielectric cavities, several workers have also adopted this figure
of merit for plasmonics~\cite{maierV,oulton,polman}. Here, I
demonstrate that the Purcell factor does not provide a quantitative
gauge for spontaneous emission enhancements in plasmonic antennas.

The spontaneous emission rate $\Gamma$ of a dipole emitter
(transition frequency $\omega$, transition dipole moment $d$) is
governed by the local density of photonic states (LDOS)
$N(\omega,\mathbf{r},\mathbf{e}_d)$ that enters Fermi's Golden
Rule~(\emph{cf.} p. 273-275 in \cite{Novotny}) through
$\Gamma(\omega,\mathbf{r},\mathbf{e}_d) = \frac{\pi d^2
\omega}{3\hbar \epsilon_0} N(\mathbf{r},\omega,\mathbf{e}_d)$. Here,
$\mathbf{r}$ is the source position and $\mathbf{e}_d$ the dipole
orientation.  In terms of the electric field Green dyadic
$\mathbb{G}(\mathbf{r},\mathbf{r'},\omega)$
 the LDOS equals~\cite{taibook,Novotny}
\begin{equation}\label{LDOS2Green}
N(\omega,\mathbf{r},\mathbf{e}_d)= \frac{6 \omega}{\pi c^2}
(\mathbf{e}_d^T \cdot
\mathrm{Im}(\mathbb{G}(\mathbf{r},\mathbf{r},\omega)) \cdot
\mathbf{e}_d).
\end{equation}
This formulation is rigorous even for dissipative  plasmonic,
systems, in which case Eq.~(\ref{LDOS2Green}) includes
quenching~\cite{Novotny,mertens}. The Purcell factor for dielectric
cavities rests on two assumptions. First, in non-dissipative
systems, normal modes can be defined, and the LDOS
(\ref{LDOS2Green}) can be rewritten as a sum over all the normal
modes
\begin{equation}
N(\omega_0,\mathbf{r},\mathbf{e}_d)= 3\sum_{\lambda,\omega}
|\mathbf{e}_d^T \cdot
\mathbf{E}_{\lambda}(\mathbf{r},\omega)|^2\delta (\omega-\omega_0).
\label{eq:LDOSsum}
\end{equation}
The second assumption (see G\'{e}rard et al~\cite{Gerard})   is that
the sum over all modes in Eq.~(\ref{eq:LDOSsum}) is dominated by
just a single term. The $\delta$-function is replaced by a
Lorentzian spectrum of width $\Delta \omega = \omega / Q$ centered
around the cavity resonance, and the spatial profile is set by the
approximate eigenfunction of the cavity, normalized to contain a
single photon within its mode volume $V$. Within this approximation,
the LDOS $N_{\mathrm{cav}} (\omega_{\rm cav},\mathbf{r}_{\rm
max},\mathbf{e}_d)$ at the spatial and spectral mode maximum,
normalized to the LDOS $N_\mathrm{host} = n\omega^2/(3\pi^2 c^3) $
of the host medium (index $n$) is well approximated by the Purcell
factor
\begin{equation}
F = \frac{3Q}{4\pi^2 V}\left(\frac{\lambda}{n}\right)^3.
\label{eq:Purcell}
\end{equation}
Maier has previously noted that reasonable numerical values for $F$
and $V$ can only be obtained for plasmonic systems if the energy
density $\epsilon |E|^2$  that enters the mode volume $V={\int
\epsilon |E|^2 d\mathbf{r}}/{\mathrm{max}(\epsilon |E|^2 )}$ is
redefined~\cite{maierV} to avoid, e.g., negative energy
densities~\cite{maierV,ruppin}. The proper energy density that is
positive and real even when $\mathrm{Re}(\epsilon)<0$,  and that
accounts for   energy stored in the material was derived by
Ruppin~\cite{maierV,ruppin}. Following Ruppins
formalism~\cite{ruppin}, the energy density used here is
$(\mathrm{Re}\epsilon + 2\omega \mathrm{Im}\epsilon/\gamma) |E|^2$,
appropriate for a Drude model
$\epsilon=1-\omega_p^2/\omega(\omega+i\gamma)$ for the dielectric
constant of silver ($\omega_p=7.9$~eV, damping rate
$\gamma=0.06$~eV).

\begin{figure}[th]
\includegraphics[width=\figwidth]{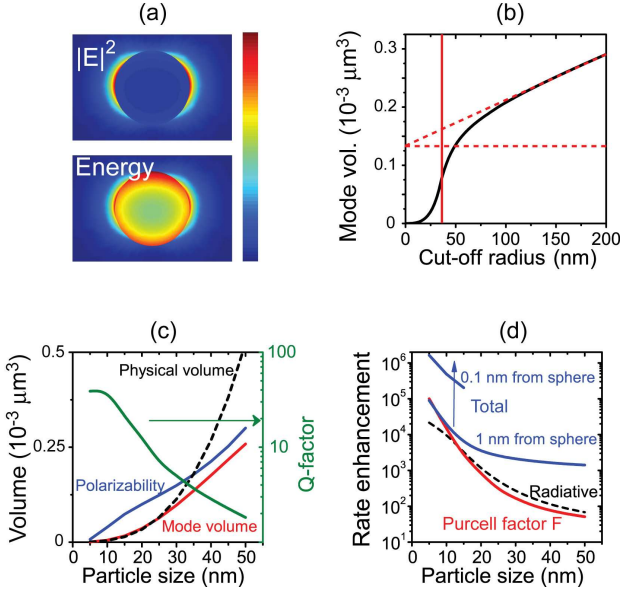}
\caption{(Color online) Contour plots of electric field $|E|^2$ and
energy density $W$ for a 35 nm radius silver sphere in vacuum (plane
wave excitation from below, horizontal polarization). The colorbar
ranges from 0 to 107 in units of the incident $|E|^2$, resp. 0 to
118 in units of $W$ of the incident wave. (b) Mode volume $\int W
d\mathbf{r}/\mathrm{max}(\epsilon(r) |E|^2)$ evaluated over a
spherical integration domain truncated at $R$, plotted versus $R$.
Vertical line: particle radius. Dashed line: linear divergence due
to radiation loss. Horizontal line: mode volume. (c) Mode volume
(red), polarizability (blue) and physical volume (dashed) versus
sphere radius for Ag spheres in vacuum. The right hand scale refers
to the quality factor $Q$ (green). (d) Red curve:
 Purcell enhancement calculated from Q and V. Dashed: exact radiative decay rate. Solid curves: total decay rate for
 a dipole at 1 nm and 0.1 nm from the sphere. \label{modevolintegral}}
\end{figure}

In this work,  the commonly employed procedure to obtain $F$
described above is compared with  exact calculations of the LDOS for
the textbook case of metal spheres. First, I consider calculation of
the mode volume. For dielectric cavities, one typically runs a
numerical solver to obtain the local field. As long as the
excitation has spatial and spectral overlap with the cavity mode, a
high $F$ ensures that the calculated spatial distribution of the
local field of the resonance is almost independent of the exact
excitation. The field distribution (total field minus excitation) is
used to evaluate the mode volume integral, with integration limits
truncated at the computation domain. Also for plasmon antennas, far
field excitation yields near-field patterns with large local field
enhancements, representative of the resonance mode profile. In this
work, the exact Green function for a sphere~\cite{taibook} is used
to calculate the local energy density using plane wave excitation.
Figure~\ref{modevolintegral}(a) shows a strongly enhanced and
confined field (enhancement relative to incident $|E|^2$ by $\sim
10^2$) just outside the sphere (sphere radius 35~nm). The energy
density reaches a maximum just inside the sphere. In
Figure~\ref{modevolintegral}(b) a plot is shown of the mode volume
integral evaluated over a truncated integration domain of radius $R$
around the plasmon sphere. Figure~\ref{modevolintegral}(b)
exemplifies a hitherto unnoted problem that is generic for all
radiatively damped resonances~\cite{pedro}, including all plasmon
antennas. The mode volume integral shows a linear divergence with
the truncation radius $R$. The divergence is understood by noting
that radiative loss signifies constant integrated far-field flux,
and hence a $1/R^2$ asymptotic fall-off of $|E|^2$. For dielectric
cavities, this divergence is generally not apparent as high Q's
imply a very small slope of the linear tail in
Fig.~\ref{modevolintegral}(b), that is often negligible in a finite
numerical computation domain. Leaving the fundamental treatment
aside~\cite{pedro}, I note that for dielectric cavities a rigorous
mode volume can be retrieved by subtracting the radiative part
(linear divergence). Figure~\ref{modevolintegral}(c) shows the
divergence-corrected mode volume for silver spheres of different
sizes. The mode volume is generally comparable to, or even smaller
(large spheres) than the physical volume, since the field energy is
mainly stored near the metal surface.

To obtain the quality factor $Q$,  Mie extinction cross
sections~\cite{taibook} are analyzed. By dividing out the
non-resonant $\omega^4$ scaling,  Lorentzian line shapes are
retrieved from which $Q$ can be extracted.
Figure~\ref{modevolintegral}(c) shows that the quality factor
decreases from $Q\sim 40$ (limited by absorption) in very small
particles to $Q\sim 3$ for larger particles. Combining $Q$ and $V$
in Fig.~\ref{modevolintegral}(d) reveals predicted Purcell factors
that range from $10^5$ for the smallest spheres to below $10^2$ for
the $50$~nm spheres.
 The key question is  in how far this
Purcell factor $F$ correctly predicts the emission rate enhancement
calculated rigorously from $\mathrm{Im}\mathbb{G}$~\cite{taibook}.
For this comparison, the maximum total decay rate enhancement almost
at the particle surface (separations shown: $1$~nm and $0.1$~nm),
where the mode field is maximum, is plotted in
Fig.~\ref{modevolintegral}(d). The Purcell factor clearly
\emph{under}-estimates the total radiative decay rate by an order of
magnitude. Strictly speaking the, Purcell factor is an estimate for
the LDOS, i.e. the total decay rate. The Purcell factor estimates
the radiative decay reasonably  for particles $>15$~nm radius.

\begin{figure}
\includegraphics[width=\figwidth]{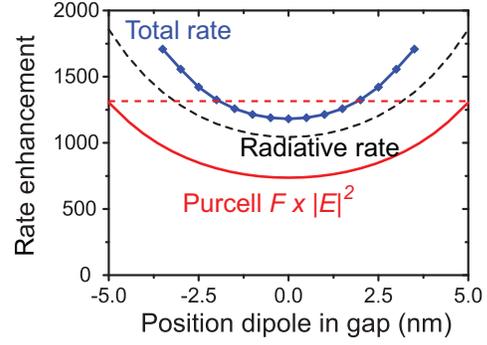}
\caption{Radiative decay rate (dashed) and total decay rate (joined
symbols) enhancement  along the centerline joining two 25 nm silver
spheres with a 10 nm gap. Enhancement is relative to the rate in the
$n=1.5$ host medium. Red: prediction from Purcell theory. The dipole
orientation is along the centerline. Right: contour plot of $|E|^2$
(scale from 0 to 1890 in units of incident $|E|^2$, dimer
illuminated by a vertically polarized plane wave).
 \label{dimer}}
\end{figure}

One might argue that single plasmon spheres set a poor example, as
they do not provide very high local field enhancements typical for
plasmonics. Figure~\ref{dimer} presents an analysis of the Purcell
factor for two spheres with a narrow gap, with a highly confined gap
mode. The calculation uses a multipole expansion multiple scattering
code provided by Garc\'{\i}a de Abajo~\cite{abajo} to analyze a
dimer of two 25 nm radius silver spheres separated by a 10 nm gap,
in a $n=1.5$ host medium. A scattering resonance with $Q=5.7$ occurs
near $\lambda =506$~nm, with a 1890 times  $|E|^2$ enhancement
polarized along the dimer axis. Purcell analysis predicts a mode
volume $V=1.32\cdot10^{-5}~\mu$m$^{3}$ and a rate enhancement
ranging from $F\sim 1300$ at a sphere surface to about 700 times
exactly in between the spheres. The rigorous calculation predicts a
rate enhancement of $\sim 1200$ times exactly in the middle of the
gap. Again the Purcell factor underestimates the actual total and
radiative decay rate enhancements by almost a factor of 2.

 Referring to the two assumptions underlying the Purcell factor, in general both assumptions will break down: Firstly, a set of normal modes
 can generally not be defined for dispersive absorbing media. Secondly, if an expansion in modes can be
 generalized (as in the case of multipole expansion for the Green function of a plasmon
 sphere) no single term is dominant,  even if the resonance is clearly due to
 one term.  For instance, at the dipolar plasmon
 sphere resonance,  the LDOS is dominated by non-resonant coupling of the emission dipole
 to higher order multipole moments of the
plasmon sphere~\cite{mertens,castanie}.  This nonresonant coupling
is not contained in a Purcell factor estimate based on $Q$ and $V$,
and corresponds to immediate quenching of emission when very close
to the metal. The  total decay rate diverges as $z^{-3}$, where $z$
is the separation between dipole and metal
surface~\cite{weber,castanie}.  Such quenching  occurs generally in
both planar and curved plasmonic structures~\cite{castanie}, and
unfortunately always exactly coincides with the field maximum where
one seeks to apply the Purcell factor. In the single sphere case,
the fact that the Purcell factor reasonably approximates the
radiative rate is consistent with the fact that most non-resonant
coupling is quenching.      For the dimer case, if the dipole is in
the middle of the gap, multipole contributions enhance both the
total and radiative decay rate while maintaining a high quantum
efficiency.

Based on the above, one is led to conclude that the Purcell factor
is not adequate for plasmon antennas. One might expect that moving
the emitter outside the regime of the extreme near field limit
allows one to retrieve the Purcell enhancement prediction, taking
into account that displacement and detuning of the emitter causes
the enhancement to decrease from $F$ in proportion to the mode field
and the Lorentzian cavity lineshape~\cite{Gerard,Novotny}. Even then
there is a fundamental difference that sets plasmon antennas aside.
Contrary to Purcell theory, the emission rate of a dipole near a
nano-antenna can follow a dispersive Fano-type frequency
dependence~\cite{mertens,feldman}. As first discussed by Dulkeith et
al. ~\cite{feldman} the lineshape can be understood as destructive
interference of radiation emitted by the emission dipole with that
radiated by the plasmon antenna. This phenomenon occurs for all
antennas where coupling to radiation is important. Incidentally, in
this regime the decay enhancement can be well understood in a point
dipole model~\cite{coevordenrmp}. The only relevant volume that
quantifies the interaction  in this case is the polarizability
(Fig.~\ref{modevolintegral}), not the mode volume.

In closing,  the Purcell factor appears  to be unsuited as a general
gauge to classify the emission enhancements of plasmonic resonances
quantitatively. While predicted Purcell factors may
 act as a qualitative design guideline, in practice one always needs to confirm by a full
calculation  that non-resonant channels are negligible.

I thank Javier Garc\'{i}a de Abajo for source code, and Ad Lagendijk
and Mischa Bonn for discussion. This work is part of the research
program of the ``Stichting voor Fundamenteel Onderzoek der Materie
(FOM),'' which is financially supported by the ``Nederlandse
Organisatie voor Wetenschappelijk Onderzoek (NWO).'' This research
is further supported by  a NWO VIDI fellowship.


\begin{thebibliography}{}

\bibitem{Novotny} L. Novotny and B. Hecht, \emph{Principles of Nano-Optics}
(Cambridge University Press, Cambridge, 2006).


\bibitem{cavityreview} K. J. Vahala, ed., \emph{Optical
Microcavities} (World Scientific, Singapore, 2004).

\bibitem{Gerard} J.-M. G\'erard and B. Gayral, %
Purcell effect for InAs quantum boxes in three-dimensional solid-state microcavities, %
 J. Lightwave Technol. \textbf{17}, 2089 (1999).



\bibitem{plasmonreview} J. A. Schuller, E. S. Barnard, W. S. Cai, Y.
C. Yun, J. S. White, and M. L. Brongersma, %
Plasmonics for extreme light concentration and manipulation, %
 Nature Mat. \textbf{9}, 193 (2010).


\bibitem{anger} P. Anger, P. Bharadwaj, and L. Novotny, %
Enhancement and quenching of single-molecule fluorescence, %
Phys. Rev. Lett. \textbf{96}, 113002 (2006).


\bibitem{sandoghdar} S. K\"{u}hn, U. H\aa kanson, L. Rogobete, and V.
Sandoghdar, %
Enhancement of single molecule fluorescence using a gold nanoparticle as an optical nano-antenna, %
Phys. Rev. Lett. \textbf{97}, 017402 (2006).

\bibitem{jaime} O.L. Muskens, V. Giannini, J.A. S\'{a}nchez Gil, and J. G\'{o}mez
Rivas, %
Strong enhancement of the radiative decay rate of emitters by single plasmonic nanoantennas, %
Nano Lett. \textbf{7}, 2871 (2007).


\bibitem{mertens} H. Mertens, A. F. Koenderink and A. Polman, %
Optimizing plasmon-enhanced luminescence near noble-metal nanospheres: comparison of the exact model and the Gersten and Nitzan approach, %
Phys. Rev. B \textbf{76}, 115123 (2007).


\bibitem{taminiau} T. H. Taminiau, F. D. Stefani, F. B. Segerink and N. F. van
Hulst, %
Optical antennas direct single-molecule emission, %
Nature Photon. \textbf{2} 234 (2008).


\bibitem{moerner} A. Kinkhabwala, Z. Yu, S. Fan, Y. Avlavesich, K.
M\"ullen, and W. E. Moerner, %
Large single-molecule fluorescence enhancements produced by a bowtie nanoantenna, %
 Nature Photon. \textbf{3}, 654 (2009).


\bibitem{lukin} A. V. Akimov, A. Mukherjee, C. L. Yu, D. E. Chang, A. S. Zibrov, P. R. Hemmer, H. Park, M. D.
Lukin, %
Generation of Single Optical Plasmons in Metallic Nanowires Coupled to Quantum Dots, %
Nature \textbf{450}, 402 (2007).


\bibitem{polman} M. Kuttge, F.J. Garc\'{i}a de Abajo and A. Polman, %
Ultrasmall Mode Volume Plasmonic Nanodisk Resonators, %
Nano Lett. \textbf{10}, 1537 (2010).

\bibitem{maierV} S. A. Maier, %
Plasmonic field enhancement and SERS in the effective mode volume picture, %
 Opt. Express  \textbf{14}, 1957 (2007).

\bibitem{oulton} R. F. Oulton, G. Bartal, D. F. P. Pile and X. Zhang, %
Confinement and propagation characteristics of subwavelength plasmonic modes, %
 New. J. Phys. \textbf{10}, 105018 (2008).


\bibitem{taibook} C.-T. Tai, \textit{Dyadic Green Functions in Electromagnetic
Theory}, 2nd ed. (IEEE, New York, 1993).


\bibitem{ruppin} R. Ruppin, %
Electromagnetic energy density in a dispersive and absorptive, %
 Phys. Lett. A \textbf{299}, 309 (2002).


\bibitem{pedro} P. de Vries, A. F. Koenderink and A. Lagendijk, to be
published.

\bibitem{abajo} F. J. Garc\'{\i}a de Abajo, %
Multiple scattering of radiation in clusters of dielectrics, %
Phys. Rev. B. \textbf{60}, 6086 (1999).


\bibitem{weber} G. W. Ford and W. H. Weber, Phys. Rep. %
Electromagnetic interactions of molecules with metal surfaces, %
\textbf{113}, 195 (1984).

\bibitem{castanie} E. Castani\'{e}, M. Boffety, and R. Carminati, %
Fluorescence quenching by a metal nanoparticle in the extreme near-field regime, %
Opt. Lett. \textbf{35}, 291 (2010).

\bibitem{coevordenrmp} P. de Vries, D.V. van Coevorden, and A.
Lagendijk, %
Point scatterers for classical waves, %
 Rev. Mod. Phys.
\textbf{70}, 447 (1998).

\bibitem{feldman} E. Dulkeith, A. C. Morteani, T. Niedereichholz, T. A. Klar,
J. Feldmann, S. A. Levi, F. C. J. M. van Veggel, D. N. Reinhoudt, M.
M\"{o}ller, and D. I. Gittins, %
Fluorescence Quenching of Dye Molecules near Gold Nanoparticles: Radiative and Nonradiative Effects, %
Phys. Rev. Lett. \textbf{89}, 203002 (2002).

\end{thebibliography}
\end{document}